\documentclass[aps,prc,preprint,groupedaddress,showpacs]{revtex4}
\usepackage{graphicx}
\begin{document}

%\preprint{}

\title{Asymptotic Freedom for Non-Relativistic Confinement}

\author{ David R. Harrington   }
%\email[drharr@physics.rutgers.edu]{Your e-mail address}
%\thanks{}
\affiliation{Deparment of Physics and Astronomy, Rutgers
University \\  136 Frelinghuysen Road \\  Piscataway, NJ
08854-8019, USA }


\date{\today}

\begin{abstract}
  Some aspects of asymptotic freedom are discussed in the context of a  simple
    two-particle non-relativistic confining potential model.  In this model
    asymptotic freedom follows from the similarity of the free-particle
    and  bound state radial wave functions at small distances and for the same angular momentum
    and the same large energy.
    This similarity, which can be understood using simple quantum
     mechanical arguments, can be used to show that the exact response function
     approaches that obtained when final state interactions are  ignored. A method of
    calculating corrections to this limit
     is given and  explicit examples are given for
    the case of the harmonic oscillator.  % insert abstract here
\end{abstract}

%\keywords{}

\maketitle

\section{Introduction}

    Asymptotic freedom was discovered to be a property of non-Abelian gauge theories nearly 30 years ago
    \cite{GrossWilczek} \cite{Politzer},
    and has been used to explain scaling in electroproduction \cite{Bjorken}. A version of asymptotic freedom also holds
    trivially in non-relativistic quantum mechanics since the Born series converges more and more rapidly
    with increasing energy.  In particular, final state interactions can be ignored in the response functions
    for inclusive scattering from condensed matter systems at large momentum transfer \cite{Sears}.  This result follows
    easily from simple quantum mechanical arguments, although the calculation of corrections can be quite
    complicated \cite{Rinat93}.  The case of confined systems is a bit more subtle, however, since the particles in the final
    state are always in discrete excited states of finite (possibly large) size and thus never really free.  The case of
    non-relativistic confinement has been discussed extensively, however, as a toy model for asymptotic freedom
    and quark-hadron duality \cite{BandG} in the structure function for electroproduction, where the production of
     resonances at
    low energies and momentum transfers averages to the smooth
    scaling curve observed at large momentum transfers.  This is
    now known to occur quite locally,  essentially for each
    resonance.  An explanation in terms of a relativistic quark
    model has been given recently \cite{IJMVanO},  while Close and Isgur
    \cite{CandI} have used a simple non-relativistic model to explain
    one aspect of duality: how  the square of a sum in the amplitude for
    production of resonances becomes approximately the sum of squares required
    for duality.  A similar non-relativistic model had been used
    earlier by Greenberg \cite{Greenberg} to investigate the
    scaling limit in terms of the Bjorken scaling variable.

        In this paper we also use a   two-body
    model to discuss non-relativistic asymptotic freedom. Such a model provides at best
     an approximate description of mesons containing heavy quarks
      \cite{Luchareview, BigiSandU}, but it is
     simple enough that some features of the large momentum transfer limit of the response function
     can be illustrated very concretely.
    In the context of the model,  asymptotic freedom follows because of a simple relation between
    the radial wave functions describing bound states in the confining potential and the
    spherical Bessel functions which describe
    free particles of the same energy and angular momentum. Not only
    do these wave functions have the same shape for high enough energy and small
    enough separation:  their relative normalizations
    are such that a form of duality follows once one
    properly allows for the comparison of a discrete sum with
    an integral.

        The paper begins with the definition of the response
    functions of the model, showing how they are related to integrals over the
    final state radial wave functions.  Both the exact case, where the final state wave functions are
    just the bound state wave functions for the discrete final states, and the approximate case using
    free-particle (plane-wave) wave functions are considered.  The close relation between the two sets of wave functions
    for the same angular momentum, similar large energies, and small radii are then described  in
    Sect. 3 for fairly general confining potentials, and the
    implications of these results for local duality
    are discussed.  A method for calculating corrections to this
    simple relation is also given, leading to a discussion  of the
    conditions under which  asymptotic freedom is a good approximation. The results are applied  in Sect. 4 to the
    case of the harmonic oscillator, where many of the
    expressions become simple analytic functions.  The
    results are summarized in Sect. 5, which also contains suggestions for further
    work in this area .

\section{Structure Functions}

    We use the  model of \cite{CandI},  consisting  of two particles of
    equal mass $m$ and reduced mass  $\mu = m/2$ carrying charges
    $e_1$ and $e_2$.  This system is  initially in its ground state in
    the spherically symmetric
    confining potential $V(r)$ and, after being hit by a "scalar
    photon", transferring momentum  $\bf{q}$ to one of the particles, it makes the transition to
    an excited state $|n_r,l,m>$  with
    energy $E_{n_r,l} $ . The probability
    for this transition is proportional to the response or structure function
    \begin{equation}
    F_{n_r,l,m}({\bf q}) = |<n_r,l,m|[e_1 e^{i\bf{r\cdot q}/2} + e_2 e^{-i\bf{r\cdot
    q}/2}] |0,0,0>|^2.
    \end{equation}
    If the polarizations $m$ of the degenerate final states  of
    a given  $n_r$ and $l$ are not measured one needs only the sum
    \begin{equation}
    F_{n_r,l} (q) = \sum_{m=-l}^{l} F_{n_r,l,m}({\bf q})
    \end{equation}
    Using completeness it is easy to see that
    \begin{equation}
    \sum_{n_r, l} F_{n_r,l}(q) = e_1^2 + e_2^2 + 2 e_1 e_2 S(q) ,
    \end{equation}
    where
    \begin{equation}
    S(q) = <0,0,0|e^{ i\bf{r\cdot q}} |0,0,0>
    \end{equation}
    is the ground state form factor.  Using standard manipulations one can write
    $F_{n_r,l}(q) $ in terms of the square of a radial integral involving the
     bound state radial wave functions $u_{n_r,l}(r)$:
    \begin{equation}
    F_{n_r,l}(q) = [e_1^2 + e_2^2 + 2 e_1 e_2 (-1)^l] (2l+1)\,
    r_{n_r,l}(q)^2 ,  \label{fnrn}
    \end{equation}
    where
    \begin{equation}
    r_{n_r,l}(q) = \int_{0}^{\infty}dr\, u_{n_r,l}(r)\, j_l (qr/2)\,
    u_{0,0}(r), \label{radintn}
    \end{equation}
    with $j_l(x) $ a spherical Bessel function.
    The $ (-1)^l $ in \ref{fnrn}  shows that the interference terms
    will tend to cancel when states of adjacent values of $l$ are
    included in a sum  \cite{CandI}.

        If the interaction of the two particles in the final
    state is completely ignored, as would follow from the
    assumption of asymptotic freedom, the final states can be
    labelled by either the relative momentum $\bf{k}$ of the two
    particles or its magnitude $k$ and the angular momentum $
    l,m$.  Both will be useful below, so we define
    \begin{equation}
    F({\bf k,q} ) =  |<{\bf k}|[e_1 e^{i{\bf r\cdot q}/2} + e_2 e^{-i{\bf r\cdot
    q}/2}] |0,0,0>|^2
    \end{equation}
    and
    \begin{equation}
     F_{l,m}(k, {\bf q} ) =  |<k,l,m|[e_1 e^{i{\bf r\cdot q}/2} + e_2 e^{-i{\bf r\cdot
    q}/2}] |0,0,0>|^2.
     \end{equation}
     With these definitions
     \begin{equation}
     F({\bf k,q} ) = e_1^2 \phi ^2(|{\bf k-q}/2|) + e_2^2 \phi
     ^2(|{\bf k+q}/2|)+ 2e_1 e_2 \phi (|{\bf k-q}/2|) \phi (|{\bf k+q}/2|)
     ,   \label{Fphi}
     \end{equation}
     where $\phi (p) $ is the ground state momentum space wave
     function which is large only when $p$ is less than $1/r_0$, where $r_0$
     indicates the size of the ground state wave function, and
    \begin{equation}
     F_{l}(k,q ) =  \sum_{m=-l}^{l} F_{l,m}(k,{\bf q} ) =[e_1^2 + e_2^2 +
      2 e_1 e_2 (-1)^l]\, (2l+1) \, r_{l}(k,q)^2 ,  \label{fkrk}
     \end{equation}
    with
      \begin{equation}
    r_{l}(k,q) =2 \int_{0}^{\infty}dr \hat{j}_{l}(kr) j_l (qr/2)
    u_{0,0}(r).  \label{radintk}
    \end{equation}
    Here, to emphasize the similarity to the bound state case, we
    have introduced the Ricatti-Bessel functions  $\hat{j}(x) =x j(x)$.  Summing $F_{l}(k,q
    )$ over $l$ produces the same structure function obtained by
    integrating $F({\bf k,q} )$ over all directions of ${\bf k}$ :
     \begin{equation}
        F(k,q) = \sum_{l=0}^{\infty} F_{l}(k,q ) = (k/2\pi )^2 \int
        d\Omega _{{\bf k}}  F({\bf k,q} ).
    \end{equation}
        The completeness relation now take the form
     \begin{equation}
     \int_ {0}^{\infty} (dk/2\pi )F(k,q) = e_1^2 + e_2^2 + 2 e_1 e_2 S(q).
    \end{equation}

    Several properties of the structure functions can be read off
    from the expressions above.  In the free case the structure
    function is large when $ {\bf k}\approx \pm {\bf q}/2 $ , and, for
    large $q $, the interference term cannot be large.  The corresponding result
    for $F_{l}(k,q) $ can be seen in Eqn. \ref{radintk}, since here the two spherical
    Bessel functions are exactly in phase only if $k \approx
    q/2$, so that $ r_{l}(k,q)$ will be maximum here and begin to decrease significantly when
    $|k-q/2|r_0 \geq 1$, where $r_0$ indicates the size of the ground state wave function.
      Using Eqn. \ref{Fphi}, a simple change of variable in the integral over the
     direction of ${\bf k}$ shows that, in the large $q$ limit, $F(k,q)$ is simply
    the  ground state probability distribution  for the component of ${\bf k}$ in
    the ${\bf q}$ direction \cite{Westscaling} . This means that, in this limit,  $F(k,q) $
    depends only on the scaling variable $k-q/2 \approx (k^2-q^2/4)/q$, which is the non-relativistic version
    of Bjorken scaling \cite{Bjorken} .

    In the confinement
    case the structure function $F_{n_r,l} $ can be large for large $q$
    only if the oscillations of
    the radial wave function $u_{n_r,l}(r)$ are in phase  with
    those of the spherical Bessel function $j_l(qr/2)$, so that asymptotic freedom will
    hold only if these radial wave functions have a form similar to that of  $ \hat{j}(kr)$. In the next section this
     requirement will be studied in more detail and it will be
     shown that the two radial wave functions have almost the
     same shapes  for the same energy and angular momentum.

\section{Radial Wave Functions}

        Assuming $V(r) \rightarrow 0$ smoothly as $r \rightarrow 0 $,
       the bound state radial wave function must have the
     same shape as the free particle radial wave functions for the same $l$ and for high
     enough energy and small enough $r$
     since they satisfy approximately the same wave equation  and both vanish at
     $r=0$.  In this section we use the WKB
     approximation to discuss the normalization of the bound
     state radial wave function and show that the result
     obtained guarantees asymptotic freedom.

        In the WKB approximation the radial wave function in the
    classically allowed region is simply
     \begin{equation}
     u_{n_r,l}(r) \approx {\cal N}_{n_r,l} \cos [\int_{r_{-}}^{r} dr' \,k_{n_r ,l} (r') - \pi /4] /\sqrt{k_{n_r ,l}
     (r)} ,
    \end{equation}
    where
    \begin{equation}
    k_{n_r ,l}(r) = \sqrt{ k_{n_r,l}^2-2\mu V(r) -(l+1/2)^2 /r^2 }
    \label{krWKB}
     \end{equation}
    is the classical radial wave number at $r$ for a system with energy
     \begin{equation}
    E_{n_r,l} = k_{n_r,l}^2/2\mu , \label{energyk}
    \end{equation}
    with the usual replacement $ l(l+1) \rightarrow (l+1/2)^2 $ \cite{Kemble}, and the classical
    turning points $ r_{-} $ and $r_{+}$ are determined by the condition $ k_{n_r ,l}(r) = 0 $.
    The normalization constant $  {\cal N}_{n_r,l} $ can  be estimated
    by assuming that, for high energy bound states, the average of
    the square of the cosine is close to $1/2$ and that the
    contributions to the normalization integral from the
    classically forbidden regions  is negligible. Normalization then requires
    \begin{equation}
    1 \approx {\cal N}_{n_r,l}^2 \int_{r_{-}}^{r_+} dr/(2 k_{n_r
    ,l}(r))
    .\label{normWKB}
    \end{equation}
    The integral is clearly proportional to the period of the
    classical motion, and it is easy to show from the Bohr-Sommerfeld condition that it
    is inversely proportional to the level splitting.  The final
    result can be written
     \begin{equation}
     {\cal N}_{n_r,l}^2 \approx 2\mu(E_{n_r +1,l} - E_{n_r
     ,l})/\pi. \label{normdelE}
    \end{equation}
    (Similar arguments can be used to derive the expression for
    the the local electron density used in  Thomas-Fermi theory \cite{HarrAJP}.)

        From the discussion above, the shape of $ u_{n_r,l}(r) $ must match that of the Ricatti-
    Bessel function $ \hat{j}_{l}(k_{n_r,l} r)$ at small $r$ .  If we assume this match
    extends into the classically allowed region and to $ k r \gg l
    $, so that the  Ricatti-Bessel function takes on its
    sinusoidal asymptotic form, it must be approximately true that
     \begin{equation}
     u_{n_r,l}(r) \approx {\cal N}_{n_r,l}\, \hat{j}_{l}(k_{n_r ,l}r)/ \sqrt{k_{n_r ,l}}
     . \label{ujhatprop}
    \end{equation}
    This relation holds only for small enough radius since the local wave number $k_{n_r ,l}(r)$ appearing in the WKB
    approximation will eventually begin to differ significantly from  from $k_{n_r ,l}$.   The radial
    integrals Eq.~(\ref{radintn},\ref{radintk}) required for calculating the structure functions,
    however, involve only values of $r$ such that the ground state
    radial wave function $ u_{0,0}(r) $  is large.  As shown in Fig. 1, this can be
    much smaller than the size of higher energy excited states
    and so it might well be possible that the radial integrals in the
    bound and free cases are essentially identical except for the
    normalization constant $ {\cal N}_{n_r,l}$.
    \begin{figure}
        \includegraphics{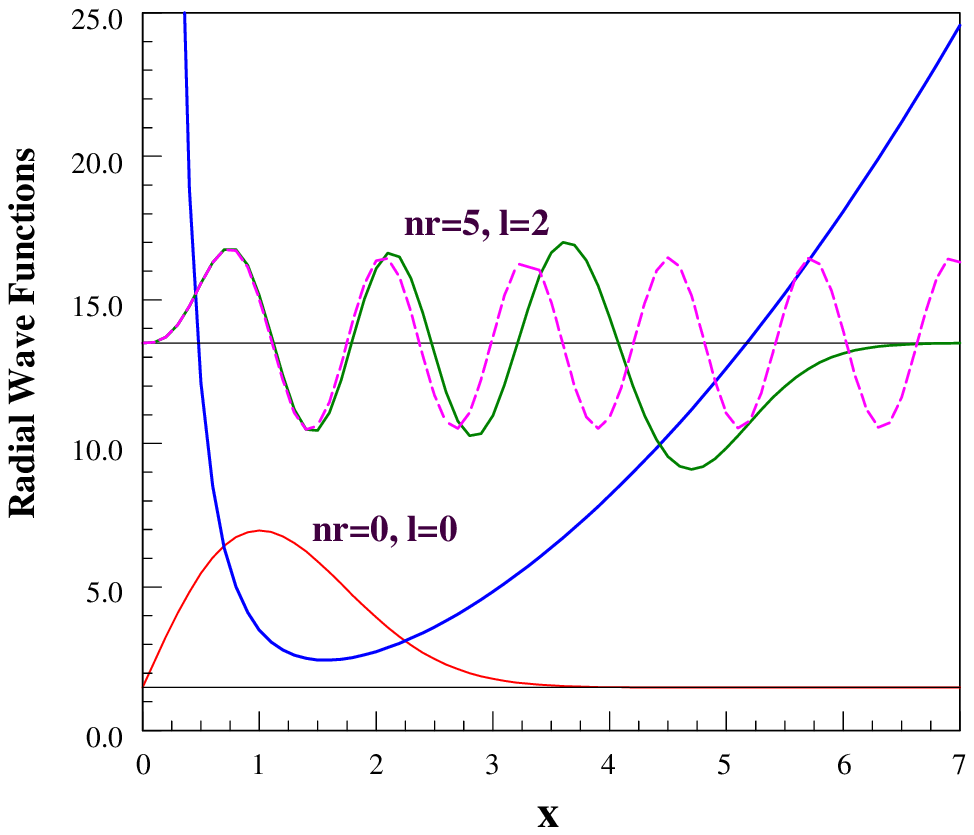}
        \caption{ Oscillator bound state radial wave function for $n_r = 5$ and $l=2$ (solid) and the free particle
        radial wave function(dashed) for the same energy and angular momentum, together with
        the ground state ($n_r=0, l=0$) radial wave function, plotted against
        the dimensionless radial variable $x = \sqrt{\alpha} r$.  The heavy curve shows the effective potential in the
        oscillator well for this angular momentum.}
    \end{figure}

        Using the relation \ref{energyk} between $E_{n_r,l} $ and $
     k_{n_r,l}$ it is easy to show that
     \begin{equation}
     \sum_{n_r} \rightarrow \int k dk/(\mu \Delta E),
    \end{equation}
    where $\Delta E$ is the energy difference appearing in Eq.
    ~\ref{normdelE}
    for energies related to $k$ by Eq. ~\ref{energyk}. Then, using (\ref{radintn}) and
    (\ref{radintk}),
    \begin{equation}
    \sum_{n_r} F_{n_r,l}(q) \approx  \int dk F_l (k,q) /(2\pi ).
    \end{equation}
    These approximate relations hold for any set  of  $n_r \,'s$
    provided  the integral on the right is over the
    corresponding range of $k's$, in the sense that the range of
    energies covered on the two sides is the same.  This of course
    is just a statement of asymptotic freedom.  Note that it holds
    for groups of states with a single definite angular momentum
    $l$, so that  the correspondence is local in angular momentum as well as energy.

        It has been stated somewhat vaguely above that the above results hold for small
    enough $r$  and high enough energy.  It would be useful to
    have a more quantitative estimate of the errors and a scheme
    for correcting them.  One method of calculating corrections has been discussed  by Rinat et al \cite{Rinat93}.
    Another, of more direct application to the approach here, using radial wave functions,
    can be obtained from the integral equation for the radial wave function \cite{Calogero}
    \begin{eqnarray}
    \hat{u}_{n_r,l}(r)& = &\hat{j}_l (k_{n_r,l}r) -(1/k_{n_r,l})
    \int_{0}^{r} dr'[\hat{j}_l (k_{n_r,l}r) \hat{n}_l (k_{n_r,l}r')
    - \hat{j}_l (k_{n_r,l}r') \hat{n}_l (k_{n_r,l}r) ]\, \nonumber \\
            &  & 2\mu V(r')\, \hat{u}_{n_r,l}(r'), \label{Calogeqn}
    \end{eqnarray}
    where $\hat{n}_l(x)$ is the Ricatti-Bessel function
     of the 2nd kind and $\hat{u}_{n_r,l}(r)$ is an un-normalized version of $u$
    which equals $\hat{j}_l (k_{n_r,l}r)$ for very small $r$.  (In
    this equation  $\hat{u}$ is normalizable only if $k_{n_r,l}$
    corresponds to an energy eigenvalue.) This integral equation
    shows clearly that the error in the proportionality between $u$
    and $\hat{j}_l$ generally increases with $r$ and the strength of $V$ ,
      but decreases as $k_{n_r,l}$ increases.
    It can be used to develop a perturbation expansion for the
    correction, the leading correction  being simply the integral term
      in (\ref{Calogeqn}) with $\hat{u}_{n_r,l}(r')$ replaced by $\hat{j}_l (k_{n_r,l}r') $.
    This expansion must, of course, diverge at large $r$, but it
    can be used to estimate the accuracy of the proportionality
    (\ref{ujhatprop}) and thus of local duality itself.   A simple change of
      variables, for example, shows that for an $r^n$ potential the leading correction
      decreases as $1/k_{n_r,l}^{n+2}$ .

\section{Harmonic Oscillator Examples}

    In the case of the harmonic oscillator potential $ V(r) =
    \frac{1}{2} K r^2 $ there are analytic expressions for most of
    the quantities discussed above.  It will be useful to define
    the usual parameters $ \omega = \sqrt{K/\mu}$ and $\alpha  =
    \mu \omega $, in terms of which $ E_{n_r, l} =\omega (2 n_r +
    l + 3/2) $.  The radial wave functions are then given by
     \begin{equation}
    u_{n_r,l}(r) = N_{n_r,l} \alpha ^{1/4} x^{l+1}
    L_{n_r}^{l+1/2}(x^2) e^{-x^2 /2},
     \end{equation}
    where  $L_{n_r}^{l+1/2}$ is an associated Laguerre polynomial,
    the dimensionless variable $x = \sqrt{\alpha} \, r $ and
    the normalization constant is
     \begin{equation}
        N_{n_r,l} = [2\, n_r !/\Gamma (n_r+l+3/2)]^{1/2} .
     \end{equation}
    Since in this case $ k_{n_r,l} = \sqrt{2 \alpha (2 n_r + l +
    3/2)}\,$ , the relation between the bound and free radial wave
    functions in terms of dimensionless quantities is
    \begin{equation}
    u_{n_r,l}(r)/\alpha ^{1/4} \approx 2 \hat{j}_l ( \sqrt{4n_r+2l+3} \, x)/[\sqrt{\pi}
    (4n_r+2l+3)^{1/4} ].
     \end{equation}
     An example of this relation is shown in Fig. 1, which
     shows that the normalized bound state and free particle radial
     wave functions for the same $l$ and energy are indeed almost identical near the
     the origin, in the region where the ground state radial wave function
     appearing in the radial integrals is large.

        For the oscillator the radial integrals have simple
     analytic expression:
     \begin{equation}
    r_{n_r,l}(q) = \sqrt{2^{n_r+l}/[n_r ! (2n_r + 2 l +1)!!]}\, \,
    (p/2)^{2 n_r + l} \, e^{-p^2 /4},
    \end{equation}
    and
       \begin{equation}
    r_{l}(k,q) =2 \sqrt{2} \kappa ( \pi/\alpha )^{1/4}
    e^{-(\kappa ^2 + p^2 ) /2}\, i_l (\kappa p),
    \end{equation}
    where $p=q/(2\sqrt{\alpha}) $ and $\kappa = k/\sqrt{\alpha} $ are dimensionless
    momenta and $i_l$ is a modified spherical Bessel function of
    the first kind.  These two functions are plotted in Fig. 2 for
    the particular case of $n_r = 5$ and $l = 2$.
     \begin{figure}
        \includegraphics{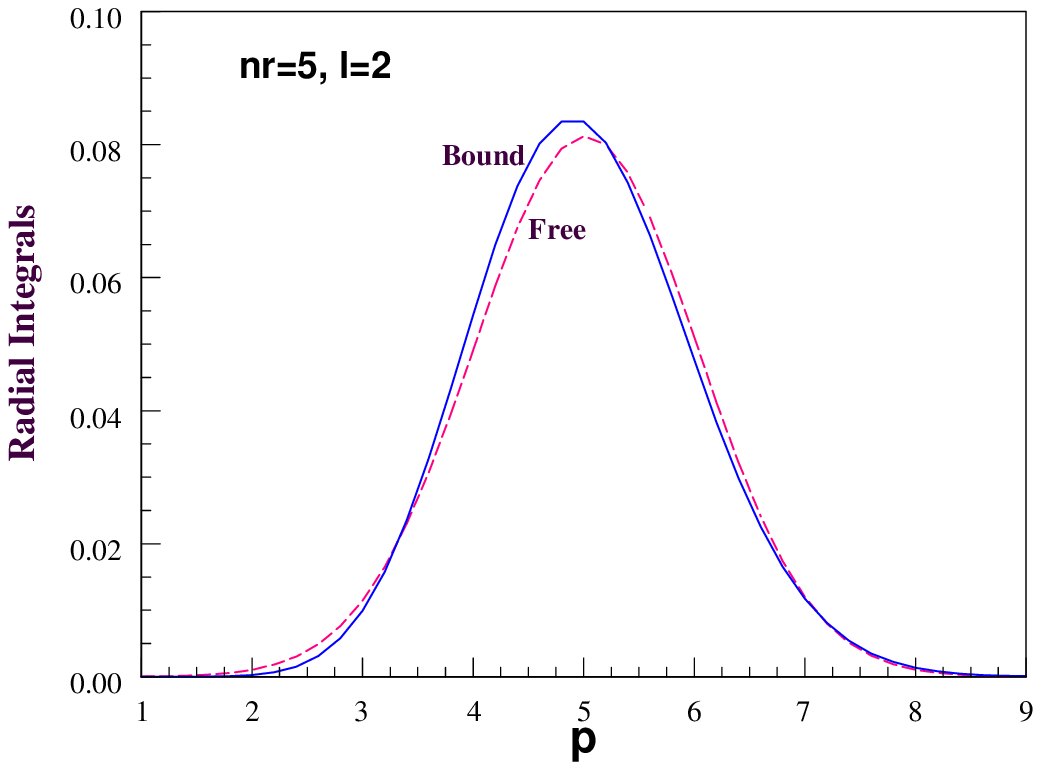}
        \caption{ Radial integrals for $n_r=5$ and $l=2$ , computed using bound state (solid) and free particle
        (dashed) radial wave functions for the same energy ($\kappa = \sqrt{2 n_r + l +3/2} = \sqrt{27} \approx 5.2$ ),
        as a function of
        the dimensionless momentum transfer  $p = q/(2 \sqrt{\alpha})$.  Note that both curves peak near  $p=5.2$.}
    \end{figure}

        With these expressions for $r_{n_r,l}(q)$ and $r_l(k,q)$ we
    can use (\ref{fnrn}) and (\ref{fkrk}) to obtain expressions for $F_{n_r,l}(q)$
    and $F_{l} (k,q)$.  These can be summed over $l$ to obtain the
    structure functions for transitions to all states at a
    definite energy:
    \begin{eqnarray}
        F_n(q) & = & \sum_{n_r,l} \delta _{n,2n_r +l} F_{n_r,l} \nonumber \\
                & = & [e_1^2 + e_2^2 + 2 e_1 e_2 (-1)^n]\, (1/n!)(p^2/2)^n e^{-p^2
                /2}\label{fnqosc}
    \end{eqnarray}
    and
    \begin{eqnarray}
        F(k,q) & = & \sum_{l=0}^{\infty} F_{l}(k,q) \nonumber \\
                & = & [e_1^2 + e_2^2 ]\, 2 \sqrt{\pi/\alpha} \,(\kappa/ p)
      [e^{-(\kappa - p)^2} - e^{-(\kappa + p)^2} ] \nonumber \\
                &  &   +  2 e_1 e_2 \,8 \sqrt{\pi /\alpha} \, \kappa^{2 }e^{-p^2 -\kappa
      ^2}. \label{fkqosc}
    \end{eqnarray}
    It should be noted that for large momentum transfers the
    $l$-dependence of the terms in the sums above becomes
    Gaussian:  $(2l+1) exp(-l^2/2n)$ in (\ref{fnqosc}) and $(2l+1)
    exp(-l^2/\kappa p) $ in (\ref{fkqosc} ).  This means that the
    number of terms which contribute significantly to the sums is
    only a few times $\sqrt{2n}$ or, equivalently, $\sqrt{\kappa p} $,
    respectively, so that the difference between the finite sum in
    (\ref{fnqosc}) and the infinite sum in (\ref{fkqosc}) is not
    significant.  This limit on the angular momenta of the
    internal states produced can be easily understood by noting
    that the impact parameters involved cannot be greater than the
    size of the initial bound state.
       The expression for $F_n (q) $ shows
    how the contributions from alternate energy levels tend to cancel for the $ e_1 e_2$ interference term.
    For $F(k,q) $, on the other hand, the interference term is always small for large values of momentum
    transfer. Furthermore, for large $p$ only the $exp[-(\kappa -p)^2] $ term can be large which,
    as will become clearer below, gives scaling of the structure function. These structure functions, of
     course, satisfy the sum rules
    \begin{equation}
        \sum_{n=0}^{\infty} F_n(q) =  \int_{0}^{\infty} dk/(2\pi ) F(k,q) = e_1^2 + e_2^2 + 2 e_1 e_2
        S(q),  \label{oscsum}
     \end{equation}
     where  here the form factor $ S(q)= e^{-p^2} $ , showing exactly how the interference term
     vanishes as $q$ increases \cite{CandI}.

        In comparing the two structure functions $F_n (q) $ and
     $F(k,q)$ the fact that in the former the energy levels are
     discrete,
     while in the latter they form a continuum, must be addressed.
     Here the $nth$ excited discrete level
     is given a half-width $ a_n$ in $\epsilon$, the energy in units of $\omega $.  This is done by
     defining
     \begin{equation}
        F(a,\epsilon ,q) = F_0 (q) \delta (\epsilon - \epsilon _0)   +  \sum_{n=1}^{\infty} F_n
        (q)\delta _{a_n} (\epsilon - \epsilon _n) ,
     \end{equation}
     where  $\epsilon _n = n + 3/2 $, and
     $\delta _a $ is a finite-width version of the Dirac delta
     function $\delta$ :
    \begin{equation}
        \delta_a (\epsilon - \epsilon _n) = (a /\pi) /[(\epsilon - \epsilon
        _n)^2 + a^2 ]  .
    \end{equation}
    In general we expect  the
        widths $a_n$ of the individual resonance peaks  to increases with $n$ .  Furthermore, if
        \begin{equation}
        F(\epsilon, q) = (\alpha /(2\pi k)) F(k,q),
    \end{equation}
      where $\epsilon $ and $k$ are related by $ \epsilon = k^2 /(2\alpha )$,
    then the integral over $\epsilon $ from zero to infinity for
    both $F(a,\epsilon, q)$ and $F(\epsilon, q)$ equals the right-hand
    expression in (\ref{oscsum}). (This equality is approximate for $F(a,\epsilon, q)$ because
    the tails of the  $F(a,\epsilon, q)$ extend into the negative $\epsilon$ region.)
     The two $F$-functions are plotted in Fig. 3 for the
    case where $ e_1 = e_2 = e$, for which the square bracket  in
    (\ref{fnqosc}) is alternately $4 e^2$ and $0$ : the contributions
    from the two particles cancel or add coherently for odd and
    even parity states respectively \cite{CandI}.
 \begin{figure}
        \includegraphics{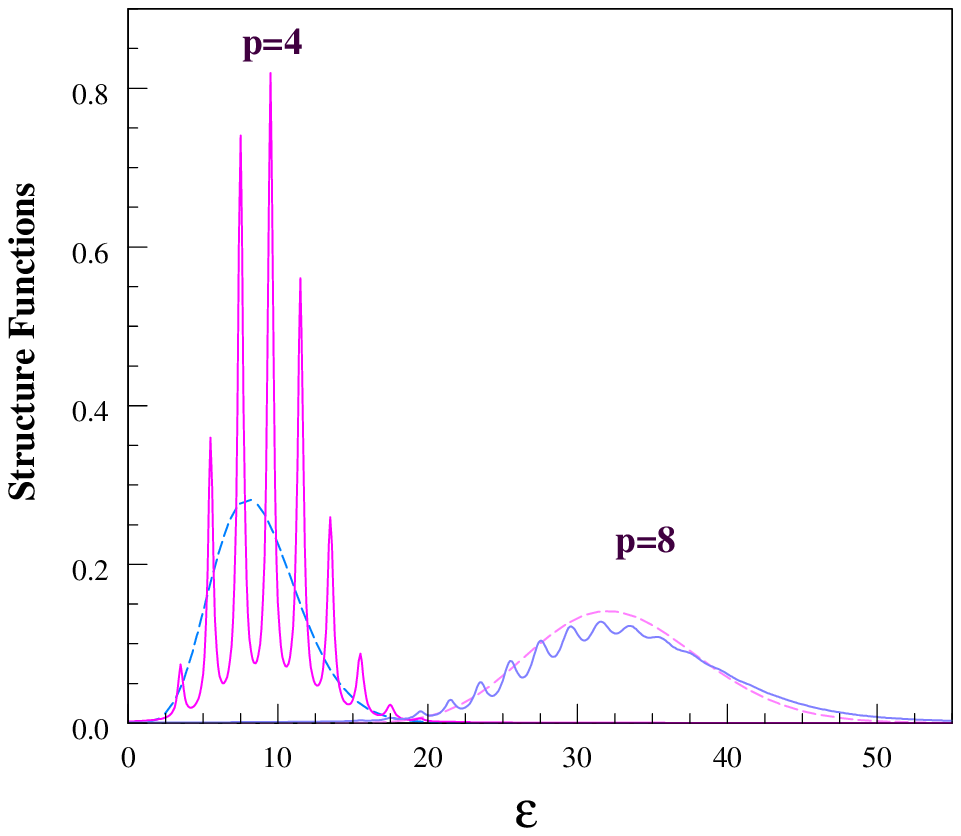}
        \caption{ Structure functions as a function of the dimensionless energy $ \epsilon = n+3/2 =k^2/2\alpha$ for
        dimensionless momentum transfers $p=q/(2\sqrt{\alpha})$ equal to 4 and 8.  The smooth curves are for
        the free case $F(\epsilon,q)$ while the curves with peaks are for the bound case $F(a,\epsilon,q)$.  In the latter cases
        the  sharp energy levels have been given  widths $a_n$
        which increase from 0.2 to 2 as their energy increases.}
    \end{figure}
     Note that the peaks of the smooth(ed) curves   appear approximately at
      $\epsilon = p^2/2$ and that the widths in $\epsilon$ increases nearly linearly
      with $p$. (The heights of the curves therefore  decrease inversely with $p$.) Thus plotting $pF$ versus
    a scaling variable $y=(\epsilon -p^2 /2)/p$ /, gives curves of a nearly Gaussian shape almost
    independent of $p$ as $p \rightarrow \infty $, as shown in Figure 4.
    This scaling variable is, aside from a trivial factor, just the $y$ variable of
    West \cite{Westscaling}, and corresponds to the component of a parton's momentum in
    the $\bf{q} $ direction before the collision.
\begin{figure}
        \includegraphics{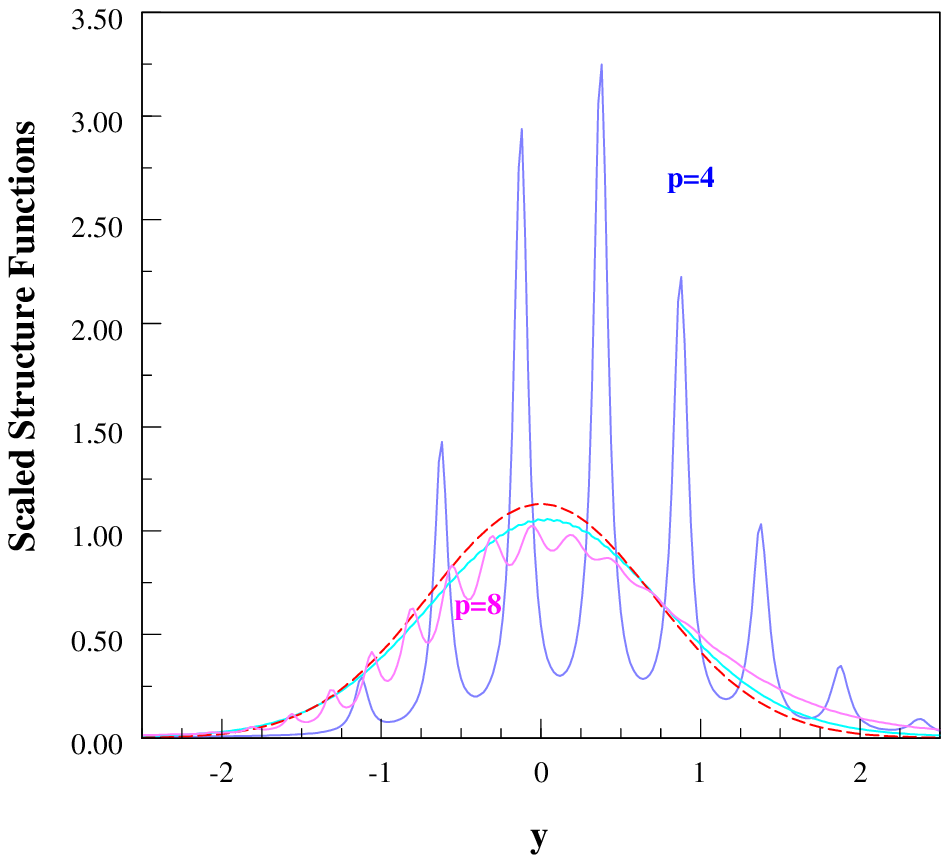}
        \caption{ Structure functions F times p as a function of the scaling variable $ y$ for
        dimensionless momentum transfers $p$ equal to 4 (large peaks), 8 (small peaks), and 32 (smooth curve).  The sharp energy levels
        have been given a width which increases from 0.2 to 2 as their energy increases. The
        dashed curve is the Gaussian limit of the scaled free particle $pF$:  $(2/\sqrt{\pi })  e^{-y^2}$. }
    \end{figure}

            We can also calculate the first order correction to the approximations in which the bound
    state wave functions
    for the oscillator are proportional to spherical Bessel functions.  The first two terms in the
    expansion can be written as
    \begin{equation}
        v_{n_r,l} (x) = v_{n_r,l}^{(0)} (x) + v_{n_r,l}^{(1)} (x),
    \end{equation}
    where
     \begin{equation}
        v_{n_r,l}^{(0)} (x) = \hat{j}_l (x)
    \end{equation}
    and
     \begin{equation}
        v_{n_r,l}^{(1)} (x) = -[1/(4 n_r + 2l +3)^2] \int_{0}^{x} dx'[\hat{j}_l (x) \hat{n}_l (x')
    - \hat{j}_l (x') \hat{n}_l (x) ]\,
             x'^2\, \hat{j}_{l}(x'),  \label{v1}
    \end{equation}
    with $x = k_{n_r,l}\,r $.
    An indication of the size of the errors and the accuracy of the leading correction above for the case
     $n_r = 5$ and $l=2$ are shown in Fig. 5.  It is clear that the corrections are small
    in the region where the ground state wave function is large, but including them
    does improve the accuracy of the expression (\ref{ujhatprop}), especially at larger values of $r$.
    Note, however, that the corrections diverge rapidly  at very large $r$.
    \begin{figure}
        \includegraphics{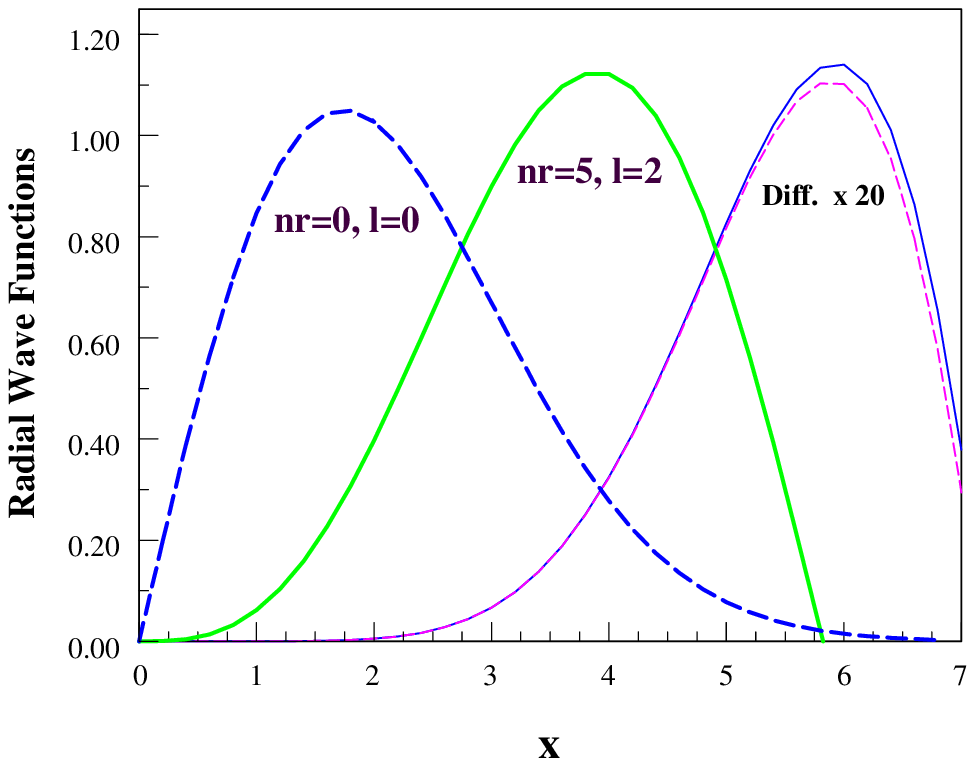}
        \caption{ Un-normalized radial wave functions and corrections for $n_r = 5$ and $l=2$ versus
         the dimensionless radius $x$.
        The heavy solid curve is the exact radial wave function, while the lighter solid and dashed curve are
         20 times the difference between $\hat{j} _2 $ and the exact wave function, and the first order estimate
        of this difference given by (\ref{v1}), respectively .  The heavy dashed curve is the ground state radial
         wave function. }
    \end{figure}

\section{Conclusion}

        In the non-relativistic case asymptotic freedom and scaling are automatic for high
    energy transfers provided the confining potential approaches
    zero smoothly as $r \rightarrow 0$. This result is almost
    trivial since the system is effectively free at small
    separations.  Only the relative normalizations of the terms in
    the sums for the bound states and the integrals in the free
    case require additional argument from the WKB approximation
    or, equivalently, the correspondence principle.

        As is well known, the resulting structure functions are
    essentially the same for  the confined and free cases,
    except for the resonance-like bumps in the former, and for
    large momentum transfers depends only on the scaling variable
    $y=(\epsilon -p^2 /2)/p  \approx  \kappa -p$, as shown in Fig.
    3.

        It would be of interest to extend these results to more
    general potentials and to semi-relativistic calculations
    \cite{Luchareview}.  For example, how would the results be modified for
    the frequently-used Coulomb-plus-linear confining potential \cite{Cornellpot}, or
    for various relativistic extensions \cite{IJMVanO, GodfreyIsgur, JVanO}?
    Another possible extension would be to many-body systems,
    including the nucleon and nuclei \cite{Westscaling}.

\section{Acknowledgement}

    The author would like to thank Professor Dieter Drechsel  and
    his colleagues in the Theory Group at the Institut f\"{u}r
    Kernphysik at the Univerit\"{a}t Mainz, where this work was done,
    for stimulating his interest
    in this topic and  for their hospitality.

\bibliography{AsympFreeRevTex4}

\end{document}